\documentclass[preprint,aps,prd,showpacs]{revtex4}
\usepackage{amssymb}%
\usepackage{graphicx}
\usepackage{subfigure}
\usepackage{amsmath}
\usepackage{bm}
\usepackage{relsize}
\RequirePackage{xspace}
\begin{document}


\title{\mbox{}\\[10pt]
QCD radiative correction to color-octet $J/\psi$ inclusive
production at B Factories}

\author{Yu-Jie Zhang $~^{(a)}$, Yan-Qing Ma $~^{(a)}$, Kai Wang $~^{(a)}$, and Kuang-Ta
Chao $~^{(a,b)}$} \affiliation{ {\footnotesize (a)~Department of
Physics and State Key Laboratory of Nuclear Physics and Technology,
Peking University,
Beijing 100871, China}\\
{\footnotesize (b)~Center for High Energy Physics, Peking
University, Beijing 100871, China}}





\begin{abstract}
In nonrelativistic Quantum Chromodynamics (NRQCD), we study the
next-to-leading order (NLO) QCD radiative correction to the
color-octet $J/\psi$ inclusive production at B Factories. Compared
with the leading-order (LO) result, the NLO QCD corrections are
found to enhance the short-distance coefficients in the color-octet
$J/\psi$ production $ e^+ e^-\to c \bar c (^3P_0^{(8)}\ {\rm or} \
^3P_0^{(8)})g$ by a factor of about 1.9. Moreover, the peak at the
endpoint in the $J/\psi$ energy distribution predicted at LO can be
smeared by the NLO corrections, but the major color-octet
contribution still comes from the large energy region of $J/\psi$.
By fitting the latest data of $\sigma(e^{+}e^{-}\to
J/\psi+X_{\mathrm{non-c\bar{c}}})$ observed by Belle, we find that
the values of color-octet matrix elements are much smaller than
expected earlier by using the naive velocity scaling rules or
extracted from fitting experimental data with LO calculations. As
the most stringent constraint by setting the color-singlet
contribution to be zero in $e^{+}e^{-}\to
J/\psi+X_{\mathrm{non-c\bar{c}}}$, we get an upper limit of the
color-octet matrix element,  $\langle 0| {\cal
O}^{J/\psi}[{}^1S_0^{(8)}]|0\rangle + 4.0\,\langle0| {\cal
O}^{J/\psi} [{}^3P_0^{(8)}]|0\rangle/m_c^2 <(2.0 \pm 0.6)\times
10^{-2}~{\rm GeV}^3$ at NLO in $\alpha_s$.

\end{abstract}

\maketitle


\section{Introduction}
It is widely believed that the heavy quarkonium production and
annihilation decay can be described by an effective theory,
non-relativistic quantum chromodynamics (NRQCD)\cite{Bodwin:1994jh}.
In the NRQCD factorization approach, the production of a heavy
quarkonium is described by a series of heavy quark pair states,
which are produced at short-distances, and then evolve into the
heavy quarkonium at long-distances by emitting or absorbing soft
gluons. The long-distance NRQCD matrix elements are scaled by the
relative velocity $v$ between quark and antiquark in the quarkonium
rest frame. And the heavy quark pair states at short-distances
should not only have the same quantum numbers as those of the
quarkonium, but also have other different quantum numbers.  In
particular, such heavy quark pairs can be in a color-octet state.
This is the so called color-octet mechanism.

The color-octet scenario seems to acquire some significant successes
in describing heavy quarkonium decay and production. But recently,
several next-to-leading order (NLO) QCD corrections for the
inclusive and exclusive heavy quarkonium production in the
color-singlet piece are found to be large and significantly relieve
the conflicts between the color-singlet model predictions and
experiments. It may imply, though inconclusively, that the
color-octet contributions in the production processes are not as big
as previously expected, and the color-octet mechanism should be
studied more carefully. Lots of work have been done to investigate
the color-octet mechanism in NRQCD  for heavy quarkonium production.

The current experimental results on $J/\psi$ photoproduction cross
sections at HERA seem to prefer the NLO color-octet predication
\cite{Butenschoen:2009zy}, rather than the  description of the NLO
color-singlet
piece~\cite{Kramer:1995nb,Chang:2009uj,Artoisenet:2009xh,Li:2009fd}.
The DELPHI data favor the NRQCD color-octet mechanism for $J/\psi$
production in $\gamma \gamma \rightarrow J/\psi X$
\cite{Abdallah:2003du,Klasen:2001cu}.

The NLO QCD corrections to $J/\psi$ and $\Upsilon$ production at the
Tevatron and LHC were calculated including the color-singlet
piece~\cite{Campbell:2007ws,Artoisenet:2007xi,Li:2008ym} and color
octet piece~\cite{Gong:2008ft}. The NLO color-singlet contributions
are found to significantly enhance  the cross sections especially in
the large $p_T$ region. The NLO QCD corrections to polarizations of
$J/\psi$ and $\Upsilon$ at the Tevatron and LHC were also calculated
~\cite{Gong:2008hk,Gong:2008sn,Gong:2008ft}. The QED contributions
to the production of $J/\psi$ there and NLO QCD corrections were
also calculated ~\cite{He:2009zzb,He:2009cq}. The NLO relativistic
corrections to $J/\psi$ production at the Tevatron and LHC were
considered too~\cite{Fan:2009zq}. The experimental data of
polarizations favor the NLO QCD corrections of the color singlet
piece rather than the color octet piece. Recent developments and
related topics in quarkonium production can be found in
Refs.~\cite{Brambilla:2004wf,Lansberg:2006dh,Lansberg:2008zm}.

The charmonium production in $e^+e^-$ annihilation at B factories
has also provided an important test ground for NRQCD and color-octet
mechanism. The large discrepancies of $J/\psi$ production via double
$c\bar c$ (including a hidden or an open charm pair) in $e^+e^-$
annihilation at B factories between LO theoretical
predictions\cite{Braaten:2002fi,
Liu:2002wq,Hagiwara:2003cw},\cite{Liu:2003zr,Liu:2003jj} and
experimental results ~\cite{Abe:2002rb,Aubert:2005tj} once were
challenging issues but now are largely resolved by higher order
corrections: NLO QCD\cite{Zhang:2005ch,Zhang:2006ay,Zhang:2008gp,
Gong:2008ce,Gong:2007db,Gong:2009ng,Gong:2009kp,Ma:2008gq} and
relativistic
\cite{He:2007te,Bodwin:2006ke,Bodwin:2007ga,Elekina:2009wt,He:2009uf}
corrections, and the results show that the color-singlet NLO
corrections (both in $\alpha_s$ and $v$) may increase the cross
section of double charmonium production e.g. $e^+e^-\to
J/\psi\eta_c$ by an order of magnitude, and indicate that the
color-singlet contributions are overwhelmingly dominant in most
cases and there seems no much room for the color-octet contributions
in charmonium production in $e^+e^-$ annihilation at B factories
(discussions on the light-cone and other approaches can be seen in
\cite{MaandSi:2004,bondar,Zhang:2008ab,Bodwin:2006dm} ).

In the $J/\psi$ inclusive production $e^+e^-\to J/\psi+X$ at B
factories, there are two color-octet processes. One is $e^+e^-\to
q\bar q+J/\psi+X (q=u,d,s)$ studied in Ref.\cite{Yuan:1996ep}, where
the light quark $q$ (or $\bar q$) emits a hard gluon which turns
into a color-octet $^3S_1$ $c\bar c$ pair fragmenting into a
$J/\psi$ with soft hadrons. But the short-distance coefficient of
this color-octet process was found to be negligible at
$\sqrt{s}=10.6$~GeV and can only be important at much higher
energies than $\sqrt{s}=10.6$~GeV\cite{Yuan:1996ep}. Therefore, this
process can be ignored at B factories.

The other color-octet process $ e^+ e^-\to c \bar c (^3P_0^{(8)}\
{\rm or} \ ^3P_0^{(8)})g$ was studied by Braaten and
Chen\cite{Braaten:1995ez}. Based on the LO NRQCD calculation, they
predicted that the $J/\psi$ production is dominated by the region
near the upper endpoint in the $J/\psi$ energy distribution, and the
width of the peak near the endpoint is of the order of 150 MeV. But
the measured $J/\psi$ spectra in $e^+e^-$ annihilation by BaBar
\cite{Aubert:2001pd} and Belle\cite{Abe:2001za} do not exhibit any
enhancement near the endpoint.

On the experimental side, the total prompt $J/\psi$ cross sections
in $e^+e^-$ annihilation were measured to be $\sigma_{tot}=2.52\pm
0.21 \pm 0.21 $~pb by BaBar \cite{Aubert:2001pd}, whereas Belle gave
a much smaller value $\sigma_{tot}=1.47\pm 0.10 \pm 0.13 $~pb
\cite{Abe:2001za}. Obviously, the large discrepancy between the two
measurements should be further clarified. On the theoretical side,
for the $J/\psi$ inclusive production cross section, the
color-singlet contributions including $e^+e^-\to J/\psi+gg$,
$e^+e^-\to J/\psi+c\bar c$ and $e^+e^-\to J/\psi+q\bar qgg(q=u,d,s)$
at LO in $\alpha_s$ were estimated to be only about $0.4 \sim
0.9$~pb \cite{Yuan:1996ep,cs}, which might imply that the
color-octet contribution should play an important role in the
inclusive $J/\psi$ production\cite{Yuan:1996ep}. However, the ratio
of $J/\psi$ production rate through double $c\bar c$ to that of
$J/\psi$ inclusive production measured by Belle\cite{Abe:2002rb}
\begin{eqnarray}
R_{c \bar c}=\frac{\sigma[e^+e^- \to J/\psi + c \bar c]}
{\sigma[e^+e^- \to J/\psi +X]} &=& 0.59 ^{+0.15}_{-0.13}\pm 0.12,
\end{eqnarray}
are much larger than LO NRQCD predictions. If only including the
color-singlet contribution at LO in $\alpha_s$, the ratio is about
$0.2 \sim 0.4$\cite{Yuan:1996ep,cs}. And a large color-octet
contribution to the $J/\psi$ inclusive production would enhance the
denominator and then further decrease this ratio. So, this became a
very puzzling issue. Some theoretical studies have been suggested in
resolving this problem. Fleming, Leibovich and Mehen use the
Soft-Collinear Effective Theory (SCET) to resum the color-octet
contribution\cite{Fleming:2003gt}. Lin and Zhu use SCET to analyze
the color-singlet contribution to $e^+e^- \to J/\psi gg$
\cite{Lin:2004eu}. Leibovich and Liu sum the leading and
next-to-leading logarithms in the color-singlet contribution to the
$J/\psi$  production cross section\cite{Leibovich:2007vr}. As a new
step, Zhang and Chao find the NLO QCD corrections to $e^+ e^- \to
J/\psi+c \bar c$\cite{Zhang:2006ay} to be large, and increase the
cross sections by a factor of about 2 (using the same matrix
elements as LO), making the ratio $R$ larger than the LO results.

Very recently, Belle reported new measurements\cite{Pakhlov:2009nj}
\begin{eqnarray}\label{X}
\sigma(e^+e^-\rightarrow J/\psi+
X)&\hspace{-0.2cm}=&\hspace{-0.2cm}(1.17\pm 0.02\pm 0.07)pb,
\end{eqnarray}
\begin{eqnarray}\label{CCBar}
\sigma(e^+e^-\rightarrow J/\psi+c\bar
c)&\hspace{-0.2cm}=&\hspace{-0.2cm}(0.74\pm 0.08^{+0.09}_{-0.08})pb,
\end{eqnarray}
\begin{eqnarray}\label{eq:nonCCBar}
\sigma(e^+e^-\rightarrow J/\psi+X_{non\  c\bar
c})&\hspace{-0.2cm}=&\hspace{-0.2cm}(0.43\pm 0.09\pm 0.09)pb.
\end{eqnarray}
The $J/\psi$ inclusive production cross section given in
Eq.(\ref{X}) is significantly smaller than that given previously by
BaBar\cite{Aubert:2001pd} and Belle\cite{Abe:2001za}. The double
charm production cross section given in Eq.(\ref{CCBar}) also
becomes smaller accordingly. The cross section of $ J/\psi +X_{non\
c\bar c}$ includes the color-singlet contribution of $ e^+ e^-\to
J/\psi + gg$ and the color-octet one of $ e^+ e^-\to c \bar c
(^3P_0^{(8)}\ {\rm or} \ ^3P_0^{(8)})g$. The color-singlet piece has
been investigated by including the NLO $O(\alpha_s)$
correction\cite{Ma:2008gq,Gong:2009kp} and $O(v^2)$ relativistic
correction\cite{He:2009uf}, of which each contributes an enhancement
factor of $1.2-1.3$ to the cross section of $ e^+ e^-\to J/\psi +
gg$. As a result, the color-singlet contribution has saturated the
observed value given in Eq.(\ref{eq:nonCCBar}), leaving little room
for the color-octet contribution.

In order to further clarify this problem, it is certainly useful to
study the color-octet process $ e^+ e^-\to c \bar c (^3P_0^{(8)}\
{\rm or} \ ^3P_0^{(8)})g$ itself and to examine the color-octet
effect of the next-to-leading order (NLO) QCD correction on the
$J/\psi$ production at B factories. In this paper we will focus on
the NLO QCD correction to the cross section and $J/\psi$ energy
distribution in $ e^+ e^-\to c \bar c (^3P_0^{(8)}\ {\rm or} \
^3P_0^{(8)})g$. The paper is organized as follows. In Section
\uppercase\expandafter{\romannumeral 2}, we will calculate the
leading order color-octet production cross sections. In Section
\uppercase\expandafter{\romannumeral 3} , we will calculate the NLO
virtual and real corrections. In section
\uppercase\expandafter{\romannumeral 4}, we will give the numerical
results and relations to the color-octet matrix elements. A summary
will be given in Section \uppercase\expandafter{\romannumeral 5}.

\section{Leading Order Calculation}

In the NRQCD factorization framework, we can write down the cross
section for the $J/\psi$ inclusive production as
\begin{eqnarray}
\label{NRQCDprod} d \sigma (e^+ e^- \to J/\psi + X) = \sum_n d
\hat{\sigma} (e^+ e^- \rightarrow c \bar{c}[n]+ X) \langle0| {\cal
O}^{J/\psi}[n]|0 \rangle \,,
\end{eqnarray}
Here $d \hat{\sigma}$ is the inclusive cross section for $c\bar{c}$
pair in a color and angular momentum state labeled by $[n] =
{}^{2S+1}L_J^{(i)}$ produced in $e^+e^-$ annihilation. And $S$, $L$,
$J$ is the spin, orbit, and total angular momentum quantum numbers
of the $c \bar c$, and $i = 1 (8)$ means that $c \bar c$ is in
color-singlet (octet) state. The short-distance coefficients are
calculable in a perturbation series in $\alpha_s$. The long-distance
matrix elements $\langle {\cal O}^{J/\psi}_n \rangle$ are the vacuum
matrix elements of four-fermion operators in
NRQCD~\cite{Bodwin:1994jh}. The long-distance matrix elements are
scaled by the relative velocity $v \ll 1$ of the $c$ and $\bar{c}$
quarks in the $c\bar c$ rest frame.

\begin{figure}
\includegraphics[width=13.5cm]{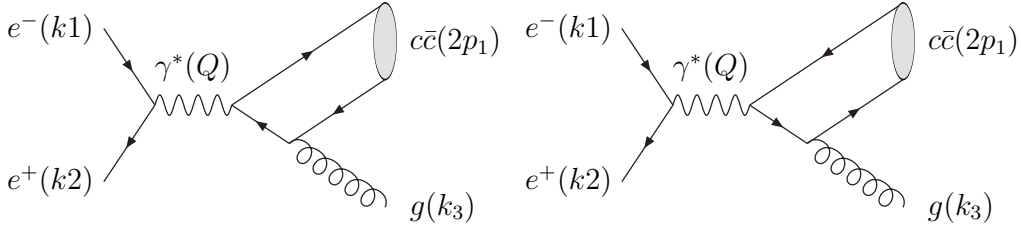}
\caption{\label{fig:LOOCT}Two Born diagrams for $e^- e^+ \to
c\bar c + g$. }
\end{figure}

At lowest order in $v$ the only term in Eq.~(\ref{NRQCDprod}) is the
color-singlet contribution, $[n]={}^3S_1^{(1)}$, which is scaled as
${\cal O}(v^3)$. The coefficient for this contribution starts at
${\cal O}(\alpha^2_s)$\cite{Yuan:1996ep,cs}, and its contribution is
away from the upper endpoint $E_{max} = (s+m_{J/\psi}^2)/ (2
\sqrt{s})$, where $s$ is the center-of-mass energy
squared\cite{Yuan:1996ep}. The color-octet contributions also start
at $O(\alpha_s^2)$, but are suppressed by $v^4 \approx 0.1$ relative
to the color-singlet contribution, and they are negligible in most
of the allowed phase-space at leading order in perturbation theory.
However, as it is pointed out in Ref.~\cite{Braaten:1995ez}, there
is an ${\cal O}(\alpha_s)$ contribution to color-octet production
near the upper endpoint, for which the Feynman diagrams are shown in
Fig.~\ref{fig:LOOCT}. Here $[n]$ is ${}^1S_0^{(8)}$ or $
^3P_J^{(8)}$, and contributions of other Fock-states are suppressed
by $v^2$. The three P-wave matrix elements are not independent at
leading order in $v^2$, and are related by
\begin{eqnarray}
\label{eq:3pjME}\langle0| {\cal O}^{J/\psi}[^3P_J^{(8)}]|0 \rangle =
(2J+1)\langle0| {\cal O}^{J/\psi}[^3P_0^{(8)}]|0 \rangle (1+{\cal
O}(v^2))
\end{eqnarray}

In the LO calculation, we refer to e.g. Ref~\cite{Braaten:1995ez}.
Momenta for the involved particles are assigned as
$e^-(k_1) e^+ (k_2)\to c\bar c (2p_1)+ g (k_3)$. In the calculation,
we use {\tt FeynArts}~\cite{feynarts} to generate Feynman diagrams
and amplitudes, {\tt FeynCalc}~\cite{Mertig:an} for the tensor
reduction, and {\tt LoopTools}~\cite{looptools} for the numerical
evaluation of the infrared (IR)-safe one-loop integrals.

$d\hat{\sigma}$ is related to the amplitude of created charm quark
pair in a color and angular momentum state $[n]$,
\begin{eqnarray}
\label{amp2}
&& \hspace{-2.4cm}{\cal A}(e^+e^- \to c\bar{c}({}^{2S+1}  L^{(8)}
_{J} )(2p_1)+g(k_3))\nonumber\\
&=& \sum_{L_{z} S_{z} }\sum_{s_1s_2}\sum_{jk}
\langle s_1;s_2\mid S S_{ z}\rangle
\langle L L_{ z };S S_{ z}\mid J J_{
z}\rangle\langle 3j;\bar{3}k\mid
8,a\rangle \times\nonumber\\
&& \left\{
\begin{array}{ll}
{\cal A}\big(e^+e^- \to
 c_j(p_1)+\bar{c}_k(p_1)+g(k_3)\big)&(L=S),\\ \left.
\epsilon^*_{\alpha}(L_Z)\frac{\partial}{\partial q^\alpha}
{\cal A}\big(e^+e^- \to
 c_j(p_1+q)+\bar{c}_k(p_1-q)+g(k_3)\big)\right|_{q=0}
&(L=P),\\
\end{array}
\right.
\end{eqnarray}
where $\langle 3j ;\bar{3}k\mid 8,a\rangle =\sqrt 2 T^a$, $\langle
s_1;s_2\mid S S_{ z}\rangle$, $\langle L L_{ z };S S_{ z}\mid J J_{
z}\rangle$ are respectively the color-SU(3), spin-SU(2), and angular
momentum Clebsch-Gordan coefficients for the $c\bar{c}$ pairs
projecting on appropriate bound states. ${\cal A}(e^+e^- \to
 c_j(p_1+q)+\bar{c}_k(p_1-q)+g(k_3))$ is the scattering
amplitude for the $c \bar c$ production. We introduce the spin
projection operators $P_{SS_z}(p,q)$ as\cite{pro}
\begin{equation}
P_{SS_z}(p,q)\equiv\sum\limits_{s_1s_2 }\langle s_1;s_2|SS_z\rangle
v(p_1-q;s_1)\bar{u}(p_1+q;s_2).
\end{equation}
Expanding $P_{SS_z}(p_1,q)$ in terms of the relative momentum $q$, we
get the projection operators at leading term and next-to-leading term
of $q$, which will be
used in our calculation, as follows
\begin{eqnarray}
\label{pjs} P_{1S_z}(p_1,0)&=&\frac{1}{\sqrt{2}}\ \epsilon\!\!
/^*(S_z)(p_1\!\!\!\!\!/+m_c). \\ \nonumber
\label{petc} P_{00}(P,0)&=&\frac{1}{\sqrt{2}}
\gamma_5(p_1\!\!\!\!\!/+m_c). \\ \nonumber
P_{1S_z}^{\alpha}(P,0)&=&\frac{1}{\sqrt{2}m_c}
[\gamma^{\alpha}\epsilon\!\!/^*(S_z)(p_1\!\!\!\!\!/+m_c)-
(p_1\!\!\!\!\!/-m_c)\epsilon\!\!/^*(S_z)\gamma^{\alpha}].
\end{eqnarray}

\section{Next-to-Leading Order Corrections}
\begin{center}
\begin{figure}
\hspace{-3.1cm}\includegraphics[width=13.2cm]{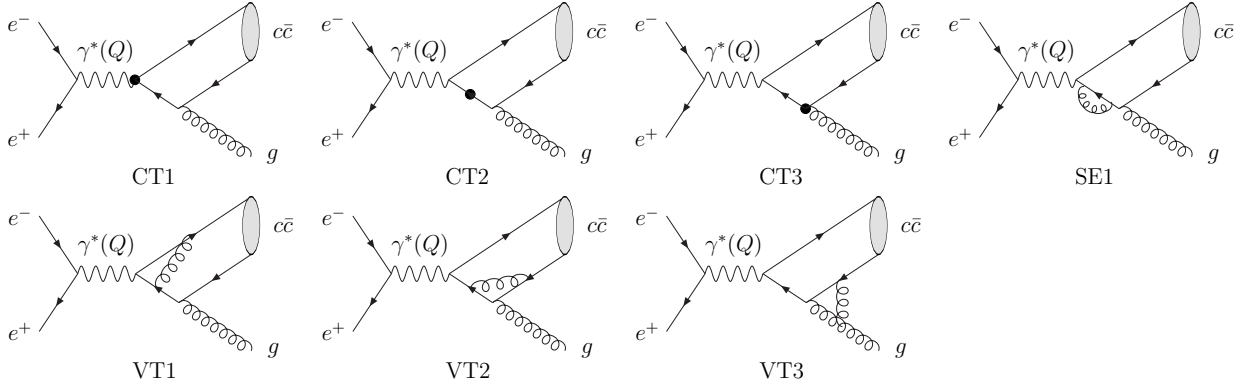}
\caption{\label{fig:ct23} Half of the counter term, self energy, and vertex
Feynman diagrams for $e^-(k_1) e^+ (k_2)\to c \bar c(2p_1) g(k_3)$.}
\end{figure}
\end{center}

\begin{center}
\begin{figure}
\includegraphics[width=12.5cm]{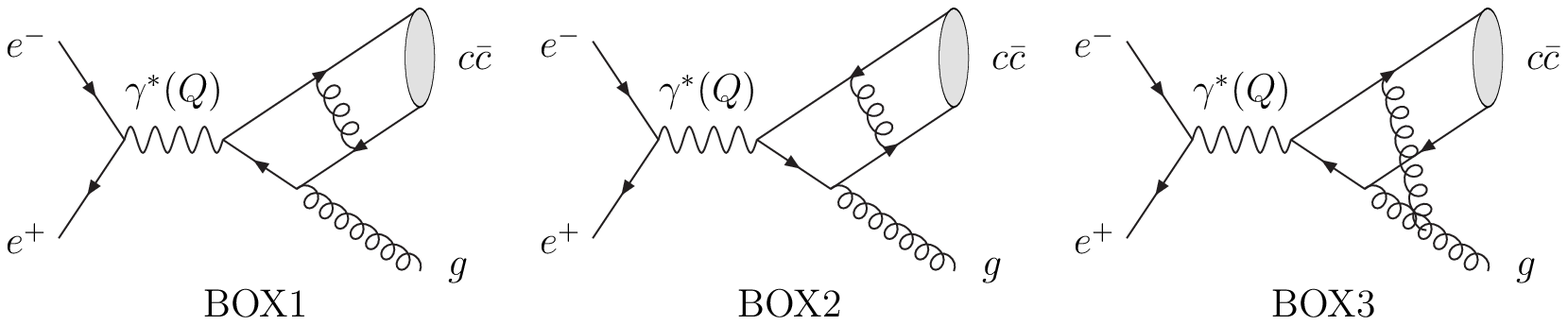}
\caption{\label{fig:box} Three box
Feynman diagrams for $e^-(k_1) e^+ (k_2)\to c \bar c(2p_1) g(k_3)$.}
\end{figure}
\end{center}

To the next-to-leading order in $\alpha_s$, the cross section is
\begin{eqnarray}
\sigma &=&\sigma_{Born} + \sigma_{virtual} + \sigma_{real}+
{\cal{O}}(\alpha^2 \alpha_s^3),
\end{eqnarray}
where
\begin{eqnarray}\label{crosssections}
 d\sigma_{Born} &=& \frac{1}{4}\ \frac{1}{2s}\ \sum
|\mathcal{M}_{Born} |^2
dPS_2 \nonumber \\
d\sigma_{virtual} &=&\frac{1}{4}\ \frac{1}{2s} \ \sum \ 2 \ {\rm
Re}(\mathcal{M}_{Born}^*\mathcal{M}_{NLO}) dPS_2\nonumber \\
d\sigma_{real} &=&\frac{1}{4}\ \frac{1}{2s}\ \sum
|\mathcal{M}_{real} |^2 dPS_3.
\end{eqnarray}
The factor $1/4$ is the average over spins of initial states. The
factor $1/2s$ is the flux factor. $\sum$ means summation over the
polarizations of initial and final states. Then we need calculate
$\mathcal{M}_{NLO}$. Here $dPS_{2(3)}$ means two(three) body phase
space. Then the self-energy and triangle diagrams are all
corresponding to propagators and vertexes of born diagrams. A half
of the counter term, self energy, and vertex Feynman diagrams for
$e^-(k_1) e^+ (k_2)\to c \bar c(2p_1) g(k_3)$ are shown in
FIG.\ref{fig:ct23}. Other diagrams can be obtained by reversing the
quark lines. And the box Feynman diagrams are shown in
FIG.\ref{fig:box}.

At NLO in $\alpha_s$, there are ultraviolet(UV), infrared(IR), and Coulomb
singularities. We choose  $D=4-2\epsilon$ dimension and relative
velocity $v$ to regularize UV, IR, and Coulomb singularities.
The self energy and vertex Feynman diagrams contain UV
singularities, which should be removed by renormalization. The
renormalization constants of  the QCD gauge coupling constant
$g_s=\sqrt{4\pi\alpha_s}$, the charm-quark mass $m$ and field
$\psi$, and the gluon field $A_\mu$ are defined as
\begin{equation}
g_s^0=Z_gg_s  ,  \qquad m^0=Z_mm  , \qquad \psi^0=\sqrt{Z_2}\psi  ,
\qquad A_\mu^0=\sqrt{Z_3}A_\mu.
\end{equation}
Here the superscript 0 means bare quantities and the renormalization
constants $Z_i=1+\delta Z_i$ with $i=g,m,2,3$. In the NLO
calculation, the precision of the quantities $\delta Z_i$ should be
${\cal O}(\alpha_s)$. We choose $Z_2$, $Z_3$, and $Z_m$ in the
on-mass-shell (OS) scheme, and $Z_g$ in the the modified
minimal-subtraction ($\overline{\rm MS}$) scheme
\begin{eqnarray}\label{renDef}
\delta Z_2^{\rm OS}&=&-C_F\frac{\alpha_s}{4\pi}
\left[\frac{1}{\epsilon_{\rm UV}}+\frac{2}{\epsilon_{\rm IR}}
-3\gamma_E+3\ln\frac{4\pi\mu^2}{m^2}+4\right]+\mathcal
{O}(\alpha_s^2),
\nonumber\\
 \delta Z_m^{\rm OS}&=&-3C_F\frac{\alpha_s}{4\pi}
\left[\frac{1}{\epsilon_{\rm UV}}-\gamma_E+\ln\frac{4\pi\mu^2}{m^2}
+\frac{4}{3}\right]+\mathcal
{O}(\alpha_s^2), \nonumber\\
 \delta Z_3^{\rm OS}&=&\frac{\alpha_s}{4\pi} (\beta_0-2C_A)
\left[\frac{1}{\epsilon_{\rm UV}} -\frac{1}{\epsilon_{\rm IR}}\right]+\mathcal
{O}(\alpha_s^2), \nonumber\\
  \delta Z_g^{\overline{\rm MS}}&=&-\frac{\beta_0}{2}\,
  \frac{\alpha_s}{4\pi}
  \left[\frac{1}{\epsilon_{\rm UV}} -\gamma_E + \ln(4\pi)
  \right]+\mathcal
{O}(\alpha_s^2),
\end{eqnarray}
where  $\mu$ is the renormalization scale, $\gamma_E$ is the Euler's
constant and $\beta_0=(11/3)C_A-(4/3)T_Fn_f$ is the one-loop
coefficient of the QCD beta function, and $n_f$ is the number of
active quark flavors. There are three massless light quarks $u,d,s$
so $n_f=3$. The charm quark $c$ is not included in the running
coupling\cite{Kramer:1995nb}. In this scheme, we do not need to
calculate the self-energy on external quark and gluon legs. Color
factors are given by $T_F=1/2,C_F=4/3,C_A=3$ in $SU(3)_c$.

Since we are calculating the NLO corrections to the LO cross
section, which is already of ${\cal O}(\alpha_s)$, we have to employ
the two-loop formula for $\alpha_s(\mu)$,
\begin{equation}
\frac{\alpha_s(\mu)}{4\pi}=\frac{1}{\beta_0L} -\frac{\beta_1\ln
L}{\beta_0^3L^2}, \label{eq:as}
\end{equation}
where $L=\ln\left(\mu^2/\Lambda_{\rm QCD}^2\right)$, and
$\beta_1=(34/3){C_A}^2-4 C_F T_F n_f-(20/3)C_A T_F n_f$ is the
two-loop coefficient of the QCD beta function.

Exchange of the longitudinal gluon between external charm quark pair
in Feynman diagram BOX1 and BOX2 of FIG.\ref{fig:box} should lead to
the Coulomb singularities $1/v$, where $v$ is the relative velocity
in the $c \bar c$ rest frame. For the Coulomb-singular color-octet
process, we find
\begin{eqnarray}
d\sigma &=& \langle {\cal O}^{J/\psi}_n \rangle_{LO}
d\hat\sigma^{(0)}\left(1- \frac{ \pi
 \alpha_s }{6 v} + \frac{\alpha_s\hat{C}}{\pi}
+{\cal{O}}(\alpha_s^2)\right). \label{eq:Coulomb}
\end{eqnarray}
The NLO color-octet matrix element $\langle {\cal O}^{J/\psi}_n
\rangle$ are proportional to $ {\pi \alpha_s /(6v)}$
\cite{Bodwin:1994jh}. It is just the Coulomb-singular term in
Eq.~(\ref{eq:Coulomb}). So that
\begin{eqnarray}
d\sigma &=& \langle {\cal O}^{J/\psi}_n \rangle_{NLO}
d\hat\sigma^{(0)}\left(1 + \frac{\alpha_s\hat{C}}{\pi}
+{\cal{O}}(\alpha_s^2)\right). \label{eq:Coulombfact}
\end{eqnarray}
Then the contribution of Coulomb singularity has to be factored out
and mapped into the color-octet matrix element $\langle {\cal
O}^{J/\psi}_n \rangle$.

The Feynman Diagrams in FIG.\ref{fig:box} and VT3 in
FIG.\ref{fig:ct23} contain IR singularity. {\tt LoopTools} can not
deal with IR divergent function and five point function, so we need
calculate it by hand. We can separate the divergence through the way
in Ref.\cite{Dittmaier:2003bc}. Then all the IR singular parts
become $C_0[-k_3,p_{1},0,0,m]$ and $C_0[p_{1},-p_{1},0,m,m]$, which
are defined as
\begin{eqnarray}
C_0[p_1,p_2,m_0,m_1,m_2]=\int \frac{\mu^{2\epsilon} \mathrm{{d}}^D
q}{(2 \pi)^D} \frac{1}{[q^2-m_0^2][(q+p1)^2-m_1^2][(q+p2)^2-m_2^2]}.
\end{eqnarray}
$C_0[p_{1},-p_{1},0,m,m]$ contains soft and Coulomb singularities,
and it will appear in Box1 and Box2 that are shown in
FIG.\ref{fig:box}. It can be regularized by $D=4-2\epsilon$
space-time dimension and relative velocity $v$:
\begin{eqnarray}\label{3pfCSvD}
C_0[p_{1},-p_{1},0,m,m]= \frac{-i}{2m^2(4\pi)^2}\left(\frac{4 \pi
\mu^2}{m^2}\right)^{\epsilon}\Gamma(1+\epsilon)\left[\,
\frac{1}{\epsilon} + \frac{\pi^2}{v} -2   + {\cal O}(\epsilon)
\right],
\end{eqnarray}
where  $v=\sqrt{-(p_{1c}-p_{1\bar c})^2}/m$. In the meson rest
frame, we have $v=|\overrightarrow{p_{1c}}- \overrightarrow{p_{1\bar
c}}|/m$, which is just the relative velocity $v$ between $c$ and
$\bar{c}$. $C_0[-k_3,p_{1},0,0,m]$ has soft and collinear
singularities, and it will appear in VT3 in FIG.\ref{fig:ct23} and
Box3 in FIG.\ref{fig:box}.
\begin{eqnarray}
C(-k_3,-k_3-p,0,0,m) &=&  \frac{ie^{-\epsilon(\gamma_E-\ln
4\pi)}}{16\pi^2} \;\;
\left(\frac{\mu^2}{m^2}\right)^{\epsilon}\;\frac{1}
{4p_1\hspace{-0.1cm}\cdot k_3} \left[ \,\frac{1}{\epsilon^2}
-\frac{2}{\epsilon}\ln\left(\frac{-2p_1\hspace{-0.1cm}\cdot
k_3}{m^2}\right)\right.\nonumber\\&&\left.+2\ln^2
\left(\frac{-2p_1\hspace{-0.1cm}\cdot k_3}{m^2}\right) {+2 {\rm
Li}_2\left(\frac{2p_1\hspace{-0.1cm}\cdot k_3+m^2}{m^2}\right)+{1
\over 2} \zeta(2)} \right].
\end{eqnarray}

\begin{figure*}
\includegraphics[width=14.5cm]{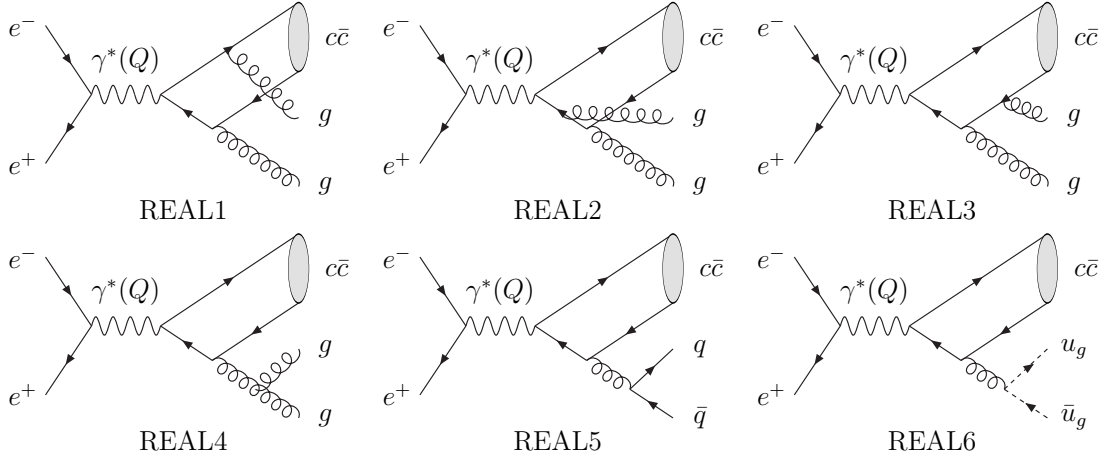}
\caption{\label{fig:real} Six of the twelve real correction
diagrams. }
\end{figure*}

There are twelve diagrams for real corrections, six diagrams for
$gg$ process, two for $q\bar q$ process, and two for ghost process.
Half of them are shown in Fig.\ref{fig:real}. The other six diagrams
can be gotten through reversing the charm quark lines. The
calculation of the real corrections is similar to the leading order
calculation, but there should appear IR singularities.
\cite{Harris:2001sx}.

\section{Numerical Result and Color-Octet Matrix Elements}
\begin{figure}
\includegraphics[width=10.0cm]{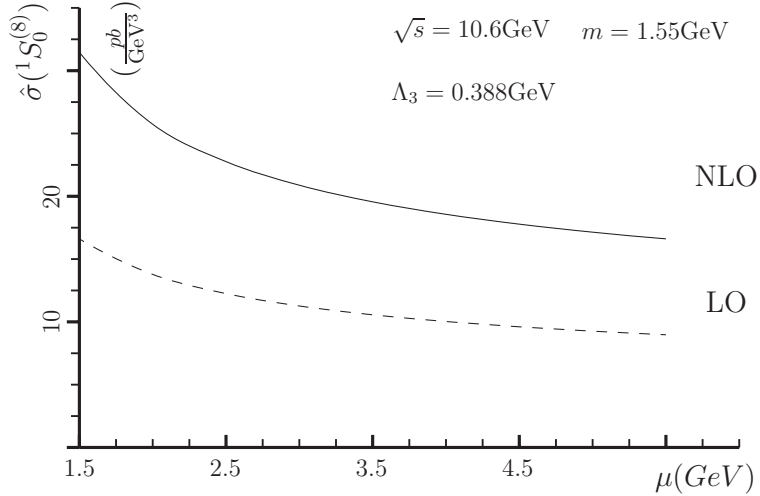}
\caption{\label{fig:1s0depmu} The short-distance coefficient $\hat
\sigma (^1S_0^{(8)})$ in  $e^+ e^-\to c \bar c (^1S_0^{(8)})g$  as
functions of the renormalization scale $\mu$. }
\end{figure}

\begin{figure}
\includegraphics[width=10.0cm]{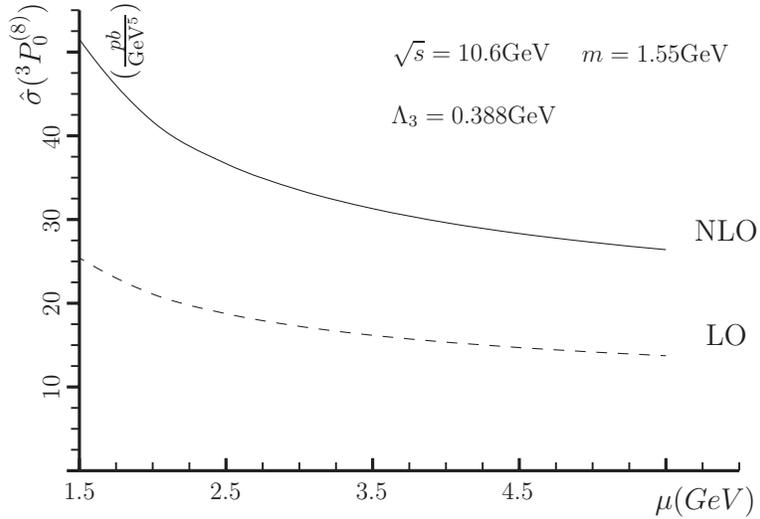}
\caption{\label{fig:3p0depmu} The short-distance coefficient $\hat
\sigma (^3P_0^{(8)})$ in  $e^+ e^-\to c \bar c (^3P_0^{(8)})g$  as
functions of the renormalization scale $\mu$. }
\end{figure}

We now turn to numerical calculations for the cross sections. Taking
$m_{J/\psi}\!=\!2m$, $m\!=\!1.55$~GeV, $\Lambda^{(3)}_{\overline{\rm
MS}}\!=\!388{\rm MeV}$, then $\alpha_s(\mu)\!=\!0.245$ for
$\mu\!=\!2m$. The cross section at LO in $\alpha_s$ is
\begin{eqnarray}
\label{eq:jpsigCOLO} \sigma(e^+ + e^-\rightarrow J/\psi
+X)&=&\Big[11 \frac{\left\langle 0\left| {\cal
O}^{J/\psi}[^1S_0^{(8)}] \right|0\right\rangle }{\rm GeV^3}+18
\frac{\left\langle 0\left| {\cal O}^{J/\psi}[^3P_0^{(8)}]
\right|0\right\rangle } {\rm GeV^5} \Big]{\rm pb};
\end{eqnarray}
while the cross section at NLO in $\alpha_s$ becomes
\begin{eqnarray}
\label{eq:jpsigCONLO} \sigma(e^+ + e^-\rightarrow J/\psi
+X)&=&\Big[21 \frac{\left\langle 0\left| {\cal
O}^{J/\psi}[^1S_0^{(8)}] \right|0\right\rangle }{\rm GeV^3}+35
\frac{\left\langle 0\left| {\cal O}^{J/\psi}[^3P_0^{(8)}]
\right|0\right\rangle } {\rm GeV^5} \Big]{\rm pb}.
\end{eqnarray}
The NLO short-distance coefficients are larger than the LO
coefficients by a factor of about $1.9$. Our LO result is consistent
with that in Ref.\cite{Braaten:1995ez}. Dependence of the
short-distance coefficient $\hat \sigma (^1S_0^{(8)})$ on the
renormalization scale $\mu$ is shown in FIG.\ref{fig:1s0depmu}, and
$\hat \sigma (^3P_0^{(8)})$ is shown in FIG.\ref{fig:3p0depmu}.

\begin{figure}
\includegraphics[width=10.0cm]{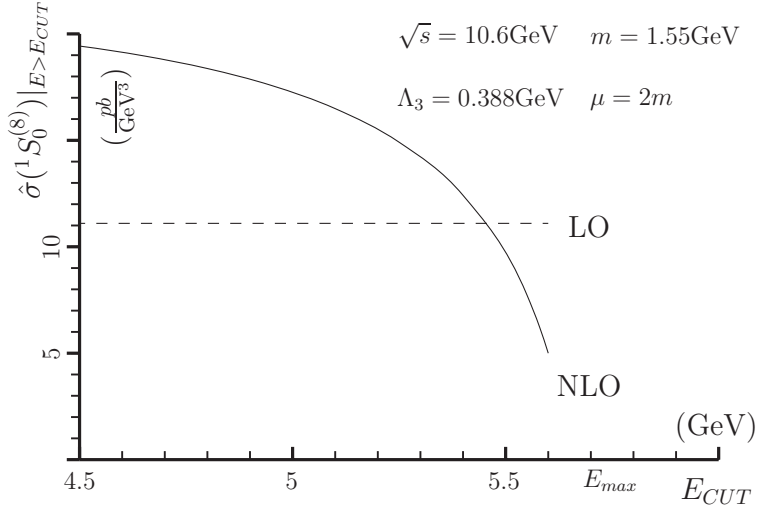}
\caption{\label{fig:1s0depEcut} The short-distance coefficient $\hat
\sigma (^1S_0^{(8)})|_{E_{J/\psi}>E_{CUT}}$ in  $e^+ e^-\to c \bar c
(^1S_0^{(8)})g$ as functions of the $c \bar c $ energy cut
$E_{CUT}$. }
\end{figure}

\begin{figure}
\includegraphics[width=10.0cm]{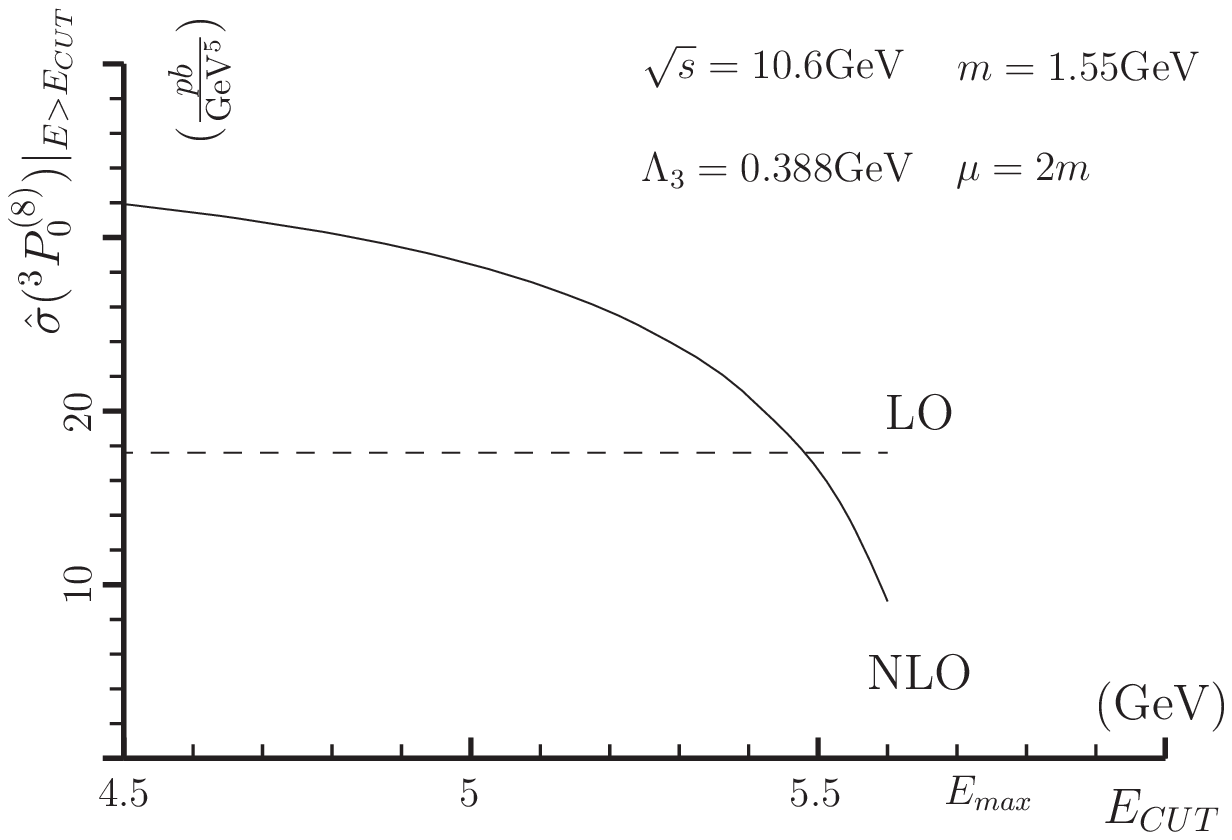}
\caption{\label{fig:3p0depEcut} The short-distance coefficient $\hat
\sigma (^3P_0^{(8)})|_{E_{J/\psi}>E_{CUT}}$ in  $e^+ e^-\to c \bar c
(^3P_0^{(8)})g$ as  functions of the $c \bar c $ energy cut
$E_{CUT}$. }
\end{figure}

If we choose an energy cut $E_{cut}$ for the $J/\psi$ in the
endpoint region, we can get the cross section
$\sigma|_{E_{J/\psi}>E_{CUT}}$. The result for $c \bar c
(^1S_0^{(8)})$ is shown in FIG. \ref{fig:1s0depEcut} and the result
for $c \bar c (^3P_J^{(8)})$ shown in FIG. \ref{fig:3p0depEcut}. We
see that at LO the short-distance coefficients  only contribute when
the $J/\psi$ energy is near the endpoint, and the NLO QCD correction
can, to some extent, smear the $J/\psi$ energy distribution near the
endpoint. Nevertheless, at NLO the most color-octet contribution
still comes from the large energy region, say $E>5$~GeV. The
evolution from the short-distance color-octet $c \bar c$ to the
final state $J/\psi$ should affect the distribution of $J/\psi$
energy via emitting or absorbing soft gluons. If we ignore this
effect, the distribution of $J/\psi$ energy in $e^+e^-$ center of
momentum frame is a delta function $\delta(E-E_{max})$ and $E_{max}
= (s+m_{J/\psi}^2)/ (2 \sqrt{s})$ at LO. Braaten and Chen analyzed
this  evolution effect, and it might broaden the energy distribution
in the order of $150$ MeV\cite{Braaten:1995ez}. But it does not
change the fact that most of the color-octet contributions
concentrate on the large energy $J/\psi$ region. Experimentally, the
observed cross sections do not exhibit any enhancement near the
endpoint\cite{Aubert:2001pd,Abe:2001za}.

The color-octet matrix element
 \begin{eqnarray}\label{eq:mk} M_k= \langle 0| {\cal
 O}^{J/\psi}[{}^1S_0^{(8)}]|0\rangle + k\,\langle0| {\cal O}^{J/\psi}
 [{}^3P_0^{(8)}]|0\rangle/m_c^2
\end{eqnarray}
was previously extracted from the Tevatron data for $J/\psi$
production with the LO calculations, which is listed in
Table.\ref{tab:COme}. If we use the minimum value of the matrix
element, $M_{3.5}=4.54 \times 10^{-2}$GeV$^3$, we can get the cross
section at LO in $\alpha_s$ from Eq.(\ref{eq:jpsigCOLO})
\begin{eqnarray}
\label{eq:jpsigCOLO2} \sigma(e^+ + e^-\rightarrow J/\psi +X)&=&(0.50
\sim 0.55){\rm pb},
\end{eqnarray}
and the cross section at NLO in $\alpha_s$  from
Eq.(\ref{eq:jpsigCONLO})
\begin{eqnarray}
\label{eq:jpsigCONLO2} \sigma(e^+ + e^-\rightarrow J/\psi
+X)&=&(0.93 \sim 1.08){\rm pb}.
\end{eqnarray}
The calculated cross section in Eq.(\ref{eq:jpsigCONLO2}) for this
color-octet process, which belongs to the $J/\psi+X_{non\  c\bar c}$
production, is much larger than the latest observed value given in
Eq.(\ref{eq:nonCCBar}): $\sigma[ J/\psi +X_{non\  c\bar c}]=0.43 \pm
0.13$~pb \cite{Pakhlov:2009nj}.
\begin{table}
\caption{\label{tab:COme}Color-octet matrix elements $M_k= \langle
0| {\cal
 O}^{J/\psi}[{}^1S_0^{(8)}]|0\rangle + k\,\langle0| {\cal O}^{J/\psi}
 [{}^3P_0^{(8)}]|0\rangle/m_c^2$}
\begin{tabular}{|c|c|c|}\hline
&\hspace{0.5cm} $k $\hspace{0.5cm} &
\hspace{0.5cm}$M_k(10^{-2}$GeV$^3)$\hspace{0.5cm}
\\ \hline
 Cho and Leibovich \cite{Cho:1995vh,Cho:1995ce}&3& $6.5 \pm 1.5$ \\
Braaten, {\it et al.}  MRST98LO\cite{Braaten:1999qk} &3.4 & $8.7
\pm 0.9$ \\
Braaten,{\it et al.}  CTEQ5L\cite{Braaten:1999qk} &3.4 & $6.6 \pm
0.7$ \\ Kramer \cite{Kramer:2001hh} &3.5& $4.54  \pm 1.11$\\ \hline
\end{tabular}
\end{table}

In fact, for the $e^+e^-\to J/\psi+X_{non\  c\bar c}$ production,
the color-singlet process $e^+e^-\to J/\psi +gg$ has been found to
make a dominant contribution to the cross section: $\sigma(e^+e^-\to
J/\psi +gg)=0.4 \sim 0.7$~pb at NLO in $\alpha_s$ and $v^2$
\cite{Ma:2008gq,Gong:2009kp,He:2009uf}, thus leaves little room to
the color-octet contributions. This gives a very stringent
constraint on the color-octet contribution, and may imply that the
values of color-octet matrix elements are much smaller than expected
earlier by using the naive velocity scaling rules or extracted from
fitting experimental data with the leading-order calculations.

Even if we disregard the dominant contribution of $e^+e^-\to J/\psi
+gg$ to $e^+e^-\to[ J/\psi +X_{non\  c\bar c}]$ by setting the
color-singlet contribution $\sigma{(e^+e^-\to J/\psi +gg)}$ to be
zero, and combining Eq.(\ref{eq:nonCCBar}) with
Eq.(\ref{eq:jpsigCONLO}), we can get an upper bound of the
color-octet matrix element:
\begin{eqnarray}\label{eq:MEupLim}
 M_{4.0}^{NLO}&<&(2.0 \pm 0.6)\times 10^{-2}~{\rm GeV}^3
\end{eqnarray}
All the values of the color-octet matrix elements listed in
Tab.\ref{tab:COme} are larger than this upper bound  by  at least a
factor of $2$ .


\section{Summary}

In summary, we find that the NLO QCD radiative corrections can
enhance the short distance coefficient of color-octet $J/\psi$
production at B factories via $ e^+ e^-\to c \bar c (^1S_0^{(8)}\
{\rm or} \ ^3P_0^{(8)})g$ with a K factor (the ratio of cross
sections of NLO to LO) of about 1.9. The NLO QCD correction smears
the $J/\psi$ energy distribution near the endpoint. But the most
color-octet contribution is still from the large energy region.  The
values of color-octet matrix elements are much smaller than expected
earlier by using the naive velocity scaling rules or extracted from
fitting the experimental data with the leading-order calculations.
If we ignore the dominant color-singlet contribution by setting the
color-singlet contribution to be zero, and use the color-octet
contribution to saturate the latest observed production cross
section $\sigma(e^+e^-\rightarrow J/\psi+X_{non\ c\bar c})$, we get
the most stringent upper bound for the color-octet matrix element:
$\langle 0| {\cal
 O}^{J/\psi}[{}^1S_0^{(8)}]|0\rangle + 4.0\,\langle0| {\cal O}^{J/\psi}
 [{}^3P_0^{(8)}]|0\rangle/m_c^2 <(2.0 \pm 0.6)\times 10^{-2}~{\rm GeV}^3$ at NLO in $\alpha_s$.

\begin{acknowledgments}
We thank G. Bodwin for useful comments. This work was supported by
the National Natural Science Foundation of China (No 10805002, No
10675003, No 10721063) and the Ministry of Science and Technology of
China (2009CB825200).

\end{acknowledgments}



\begin{thebibliography}{}


\bibitem{Bodwin:1994jh}
  G.~T.~Bodwin, E.~Braaten, and G.~P.~Lepage,
  Phys.\ Rev.\  D {\bf 51}, 1125 (1995)
  [Erratum-ibid.\  D {\bf 55}, 5853 (1997)]
  [arXiv:hep-ph/9407339].




\bibitem{Butenschoen:2009zy}
  M.~Butenschoen and B.~A.~Kniehl,
  arXiv:0909.2798 [hep-ph].







\bibitem{Kramer:1995nb}
  M.~Kramer,
  Nucl.\ Phys.\  B {\bf 459}, 3 (1996)
  [arXiv:hep-ph/9508409].


\bibitem{Chang:2009uj}
  C.~H.~Chang, R.~Li and J.~X.~Wang,
  arXiv:0901.4749 [hep-ph].


\bibitem{Artoisenet:2009xh}
  P.~Artoisenet, J.~M.~Campbell, F.~Maltoni and F.~Tramontano,
  arXiv:0901.4352 [hep-ph].


\bibitem{Li:2009fd}
  R.~Li and K.~T.~Chao,
  Phys. Rev. D79, 114020 (2009) [arXiv:0904.1643 [hep-ph]].




\bibitem{Klasen:2001cu}
  M.~Klasen, B.~A.~Kniehl, L.~N.~Mihaila and M.~Steinhauser,
  Phys.\ Rev.\ Lett.\  {\bf 89}, 032001 (2002)
  [arXiv:hep-ph/0112259].

\bibitem{Abdallah:2003du}
  J.~Abdallah {\it et al.}  [DELPHI Collaboration],
  Phys.\ Lett.\  B {\bf 565}, 76 (2003)
  [arXiv:hep-ex/0307049].













\bibitem{Artoisenet:2007xi}
  P.~Artoisenet, J.~P.~Lansberg and F.~Maltoni,
  Phys.\ Lett.\  B {\bf 653}, 60 (2007)
  [arXiv:hep-ph/0703129].


\bibitem{Campbell:2007ws}
  J.~Campbell, F.~Maltoni and F.~Tramontano,
  Phys.\ Rev.\ Lett.\  {\bf 98}, 252002 (2007)
  [arXiv:hep-ph/0703113].

\bibitem{Li:2008ym}
  R.~Li and J.~X.~Wang,
  Phys.\ Lett.\  B {\bf 672}, 51 (2009)
  [arXiv:0811.0963 [hep-ph]].





\bibitem{Gong:2008ft}
  B.~Gong, X.~Q.~Li and J.~X.~Wang,
  arXiv:0805.4751 [hep-ph].

\bibitem{Gong:2008hk}
  B.~Gong and J.~X.~Wang,
  arXiv:0805.2469 [hep-ph].

\bibitem{Gong:2008sn}
  B.~Gong and J.~X.~Wang,
  arXiv:0802.3727 [hep-ph].






\bibitem{He:2009zzb}
  Z.~G.~He, R.~Li and J.~X.~Wang,
  arXiv:0904.2069 [hep-ph].


\bibitem{He:2009cq}
  Z.~G.~He, R.~Li and J.~X.~Wang,
  arXiv:0904.1477 [hep-ph].




\bibitem{Fan:2009zq}
  Y.~Fan, Y.~Q.~Ma and K.~T.~Chao,
  arXiv:0904.4025 [hep-ph].




\bibitem{Brambilla:2004wf}
N.~Brambilla {\it et al.},
arXiv:hep-ph/0412158.

\bibitem{Lansberg:2006dh}
  J.~P.~Lansberg,
  Int.\ J.\ Mod.\ Phys.\  A {\bf 21}, 3857 (2006)
  [arXiv:hep-ph/0602091].


\bibitem{Lansberg:2008zm}
  J.~P.~Lansberg {\it et al.},
  arXiv:0807.3666 [hep-ph].




\bibitem{Braaten:2002fi}
  E.~Braaten and J.~Lee,
  Phys.\ Rev.\  D {\bf 67}, 054007 (2003)
  [Erratum-ibid.\  D {\bf 72}, 099901 (2005)]
  [arXiv:hep-ph/0211085].


\bibitem{Liu:2002wq}
  K.~Y.~Liu, Z.~G.~He and K.~T.~Chao,
  Phys.\ Lett.\  B {\bf 557}, 45 (2003)
  [arXiv:hep-ph/0211181].
  K.~Y.~Liu, Z.~G.~He and K.~T.~Chao,
  Phys.\ Rev.\  D {\bf 77}, 014002 (2008).

\bibitem{Hagiwara:2003cw}
  K.~Hagiwara, E.~Kou and C.~F.~Qiao,
  Phys.\ Lett.\  B {\bf 570}, 39 (2003)
  [arXiv:hep-ph/0305102].






\bibitem{Liu:2003zr}
  K.~Y.~Liu, Z.~G.~He and K.~T.~Chao,
  Phys.\ Rev.\ D {\bf 68}, 031501(R) (2003)
  [arXiv:hep-ph/0305084].

\bibitem{Liu:2003jj}
  K.~Y.~Liu, Z.~G.~He and K.~T.~Chao,
  Phys.\ Rev.\ D {\bf 69}, 094027 (2004)
  [arXiv:hep-ph/0301218].



\bibitem{Abe:2002rb}
K.~Abe {\it et al.} [BELLE Collaboration],
Phys.\ Rev.\ Lett.\ {\bf 89}, 142001 (2002).
 [arXiv:hep-ex/0205104].


\bibitem{Aubert:2005tj}
  B.~Aubert {\it et al.}  [BABAR Collaboration],
  Phys.\ Rev.\ D {\bf 72}, 031101 (2005)
  [arXiv:hep-ex/0506062].

\bibitem{Zhang:2005ch}
  Y.~J.~Zhang, Y.~J.~Gao and K.~T.~Chao,
  Phys.\ Rev.\ Lett.\  {\bf 96}, 092001 (2006)
  [arXiv:hep-ph/0506076].

\bibitem{Gong:2007db}
  B.~Gong and J.~X.~Wang,
  Phys.\ Rev.\  D {\bf 77}, 054028 (2008)
  [arXiv:0712.4220 [hep-ph]].



\bibitem{Zhang:2008gp}
  Y.~J.~Zhang, Y.~Q.~Ma and K.~T.~Chao,
  Phys. Rev. D78, 054006 (2008) [arXiv:0802.3655 [hep-ph]].


\bibitem{Zhang:2006ay}
  Y.~J.~Zhang and K.~T.~Chao,
  Phys.\ Rev.\ Lett.\  {\bf 98}, 092003 (2007)
  [arXiv:hep-ph/0611086].

\bibitem{Gong:2009ng}
  B.~Gong and J.~X.~Wang,
  arXiv:0904.1103 [hep-ph].


\bibitem{Gong:2008ce}
  B.~Gong and J.~X.~Wang,
  Phys.\ Rev.\ Lett.\  {\bf 100}, 181803 (2008)
  [arXiv:0801.0648 [hep-ph]].






\bibitem{Ma:2008gq}
  Y.~Q.~Ma, Y.~J.~Zhang and K.~T.~Chao,
  Phys.\ Rev.\ Lett.\  {\bf 102}, 162002 (2009)
  [arXiv:0812.5106 [hep-ph]].



\bibitem{Gong:2009kp}
  B.~Gong and J.~X.~Wang,
  Phys.\ Rev.\ Lett.\  {\bf 102}, 162003 (2009)
  [arXiv:0901.0117 [hep-ph]].
















\bibitem{He:2007te}
  Z.~G.~He, Y.~Fan and K.~T.~Chao,
  Phys.\ Rev.\  D {\bf 75}, 074011 (2007)
  [arXiv:hep-ph/0702239].

\bibitem{Bodwin:2007ga}
  G.~T.~Bodwin, J.~Lee and C.~Yu,
  Phys.\ Rev.\  D {\bf 77}, 094018 (2008)
  [arXiv:0710.0995 [hep-ph]].

\bibitem{Bodwin:2006ke}
  G.~T.~Bodwin, D.~Kang, T.~Kim, J.~Lee and C.~Yu,
  AIP Conf.\ Proc.\  {\bf 892}, 315 (2007)
  [arXiv:hep-ph/0611002].


\bibitem{Elekina:2009wt}
  E.~N.~Elekina and A.~P.~Martynenko,
  arXiv:0910.0394 [hep-ph].

\bibitem{He:2009uf}
  Z.~G.~He, Y.~Fan and K.~T.~Chao,
  arXiv:0910.3636 [hep-ph].





\bibitem{MaandSi:2004}
J.P. Ma and Z.G. Si,
Phys. Rev. D {\bf 70}, 074007(2004); Phys. Lett. B {\bf 647},419
(2007).
%

\bibitem{bondar}
  V.~V.~Braguta, A.~K.~Likhoded and A.~V.~Luchinsky,
  Phys.\ Rev.\  D {\bf 78}, 074032 (2008)
  [arXiv:0808.2118 [hep-ph]];
  V.~V.~Braguta,
  Phys.\ Rev.\  D {\bf 78}, 054025 (2008)
  [arXiv:0712.1475 [hep-ph]];
A.E. Bondar and V.L. Chernyak, Phys. Lett. B {\bf 612}, 215 (2005);
V.V.~Braguta, A.K.~Likhoded and A.V.~Luchinsky,
Phys.\ Rev.\ D {\bf 74}, 094004 (2006);
Phys.\ Rev.\ D {\bf 72}, 074019 (2005);
D. Ebert and A.P. Martynenko, Phys. Rev. D74, 054008 (2006); H.-M.
Choi and C.-R. Ji, Phys. Rev. D76, 094010 (2007).





\bibitem{Zhang:2008ab}
  Y.~J.~Zhang, Q.~Zhao and C.~F.~Qiao,
  Phys.\ Rev.\  D {\bf 78}, 054014 (2008)
  [arXiv:0806.3140 [hep-ph]],
  X.~H.~Guo, H.~W.~Ke, X.~Q.~Li and X.~H.~Wu,
  arXiv:0804.0949 [hep-ph].
  H.~M.~Choi and C.~R.~Ji,
  Phys.\ Rev.\  D {\bf 76}, 094010 (2007)
  [arXiv:0707.1173 [hep-ph]].
  Y.~Jia,
  Phys.\ Rev.\  D {\bf 76}, 074007 (2007)
  [arXiv:0706.3685 [hep-ph]].

\bibitem{Bodwin:2006dm}
  G.~T.~Bodwin, D.~Kang and J.~Lee,
  Phys.\ Rev.\  D {\bf 74}, 114028 (2006)
  [arXiv:hep-ph/0603185].










\bibitem{Yuan:1996ep}
  F.~Yuan, C.~F.~Qiao and K.~T.~Chao,
  Phys.\ Rev.\ D {\bf 56}, 321 (1997)
  [arXiv:hep-ph/9703438].



\bibitem{Braaten:1995ez}
  E.~Braaten and Y.~Q.~Chen,
  Phys.\ Rev.\ Lett.\  {\bf 76}, 730 (1996)
  [arXiv:hep-ph/9508373].


\bibitem{Aubert:2001pd}
  B.~Aubert {\it et al.}  [BABAR Collaboration],
  Phys.\ Rev.\ Lett.\  {\bf 87}, 162002 (2001)
  [arXiv:hep-ex/0106044].



\bibitem{Abe:2001za}
  K.~Abe {\it et al.}  [BELLE Collaboration],
  Phys.\ Rev.\ Lett.\  {\bf 88}, 052001 (2002)
  [arXiv:hep-ex/0110012].




%

%

%
%
%
%


\bibitem{cs}
P. Cho and A.K. Leibovich, Phys. Rev.  D {\bf 54}, 6690 (1996); F.
Yuan, C.F. Qiao, and K.T. Chao, Phys. Rev.  D {\bf 56}, 1663 (1997);
S. Baek, P. Ko, J. Lee, and H.S. Song, J. Korean Phys. Soc. {\bf
33}, 97 (1998); V.V. Kiselev {\it et al.}, Phys. Lett.  B {\bf332},
411 (1994);
S.J.~Brodsky, A.S.~Goldhaber and J.~Lee,
Phys.\ Rev.\ Lett.\ {\bf 91}, 112001 (2003).

\bibitem{Fleming:2003gt}
  S.~Fleming, A.~K.~Leibovich and T.~Mehen,
  Phys.\ Rev.\  D {\bf 68}, 094011 (2003)
  [arXiv:hep-ph/0306139].


\bibitem{Lin:2004eu}
  Z.~H.~Lin and G.~h.~Zhu,
  Phys.\ Lett.\  B {\bf 597}, 382 (2004)
  [arXiv:hep-ph/0406121].


\bibitem{Leibovich:2007vr}
  A.~K.~Leibovich and X.~Liu,
  Phys.\ Rev.\  D {\bf 76}, 034005 (2007)
  [arXiv:0705.3230 [hep-ph]].



%
%
%
%
%
%
%
%
%
%
%
%
%
%
%
%
%
%
%
%
%
%
%
%



\bibitem{Pakhlov:2009nj}
  P.~Pakhlov {\it et al.}  [Belle Collaboration],
  Phys.\ Rev.\  D {\bf 79}, 071101 (2009)
  [arXiv:0901.2775 [hep-ex]].



\bibitem{feynarts} M. B\"ohm, A. Denner, J. K\"ublbeck,
Comput.\ Phys.\ Commun.\ {\bf 60 }(1990) 165;
  T.~Hahn,
  Comput.\ Phys.\ Commun.\  {\bf 140}, 418 (2001).




\bibitem{Mertig:an} R. Mertig, M. B\"ohm, A. Denner,
Comput.\ Phys.\ Commun.\  {\bf64 }(1991) 345.


\bibitem{looptools}
  T.~Hahn and M.~Perez-Victoria,
  Comput.\ Phys.\ Commun.\  {\bf 118}, 153 (1999).






\bibitem{pro}
  P.~Ko, J.~Lee and H.~S.~Song,
  Phys.\ Rev.\ D {\bf 54}, 4312 (1996)
  [Erratum-ibid.\ D {\bf 60}, 119902 (1999)]
  [arXiv:hep-ph/9602223].

J.H. K$\ddot{\rm u}$hn, J. Kaolan and E.G.O. Safiani, Nucl. Phys.
{\bf B157}, 125 (1979);

 B. Guberina, J.H. K$\ddot{\rm
u}$hn, R.D. Peccei and R. R$\ddot{\rm u}$ckl, Nucl. Phys. {\bf
B174}, 317 (1980).




\bibitem{Dittmaier:2003bc}
  S.~Dittmaier,
  Nucl.\ Phys.\ B {\bf 675}, 447 (2003).


\bibitem{Harris:2001sx}
  B.~W.~Harris and J.~F.~Owens,
  Phys.\ Rev.\ D {\bf 65}, 094032 (2002)
  [arXiv:hep-ph/0102128].

%



\bibitem{Cho:1995vh}
  P.~L.~Cho and A.~K.~Leibovich,
  Phys.\ Rev.\  D {\bf 53}, 150 (1996)
  [arXiv:hep-ph/9505329].




\bibitem{Cho:1995ce}
  P.~L.~Cho and A.~K.~Leibovich,
  Phys.\ Rev.\  D {\bf 53}, 6203 (1996)
  [arXiv:hep-ph/9511315].

\bibitem{Braaten:1999qk}
  E.~Braaten, B.~A.~Kniehl and J.~Lee,
  Phys.\ Rev.\  D {\bf 62}, 094005 (2000)
  [arXiv:hep-ph/9911436].

\bibitem{Kramer:2001hh}
  M.~Kramer,
  Prog.\ Part.\ Nucl.\ Phys.\  {\bf 47}, 141 (2001)
  [arXiv:hep-ph/0106120].



\end{thebibliography}
\end{document}